\documentclass[twocolumn,prl,showpacs]{revtex4}
\usepackage{amsmath}
\usepackage{amssymb}
\usepackage{graphicx}
\usepackage{bm}

\begin{document}

\title{Quantum coherence induced second plateau in high-sideband generation}

\author{J. A. Crosse}
\author{Ren-Bao Liu}
\email{rbliu@phy.cuhk.edu.hk}
\affiliation{Department of Physics and Centre for Quantum Coherence, The Chinese University of Hong Kong, Shatin, N.T., Hong Kong, China.}

\date{\today}

\begin{abstract}
Optically excited electron-hole pairs, driven by a strong terahertz (THz) field, create high-sidebands in the optical spectrum. The sideband spectrum exhibits a `plateau' up to a cutoff of $3.17 U_{p}$, where $U_{p}$ is the ponderomotive energy. This cutoff is determined, semi-classically, from the maximum kinetic energy an electron-hole pair can gain from the THz field along a closed trajectory. A full quantum treatment reveals a second, classically forbidden, plateau with a cutoff of $8U_{p}$, the maximum kinetic energy an electron-hole pair can gain from the THz field along an open trajectory. The second plateau appears because a spatially separated electron and hole can still recombine if the classical excursion is less than the width of the electron-hole wavefunction, which occurs when the coherence time is longer than the excursion time (half the THz field period). This effect broadens the range of materials and excitation conditions where high-sideband generations can occur, thereby providing a wealth of novel systems for ultrafast electro-optical applications.
\end{abstract}

\pacs{42.65.Ky, 78.20.-e, 78.20.Bh} 

\maketitle
%%%%%%%%%%%%%%%%%%%%%%%%%%%%%%%%%%%%%%%%%%%%%%%%%%%%%%%%%%%%%%%%%%%%%%

High-order harmonic generation (HHG)\cite{textbook}, which occurs when an intense laser pulse interacts with an atom, is not only of fundamental interest, as an example of extreme nonlinear optics, but also of technological interest, as a method of molecular imaging \cite{image1, image2} and soft X-ray generation \cite{xray1, xray2, xray3} - the latter of which has a wide range of biological and biomedical applications. 

HHG can be understood in terms of a three step model. In the first step an electron is ionized from the atom by an intense field. In the second, the ionised electron is accelerated by the field along a closed trajectory, returning to the ion at a later time with a higher energy. Finally, the electron recombines with the ion releasing a photon with a frequency that is an odd harmonic of the original field. The maximum energy the electron can acquire along these closed trajectories results in a plateau in the HHG spectrum with a cutoff frequency of $I_{p} + 3.17 U_{p}$ (where $I_{p}$ is the atomic binding energy and $U_{p}=e^2E^2/4m_{e}\omega^2$ is the ponderomotive energy)\cite{kulander, corkum, lewenstein}.

\begin{figure}
\centering
\includegraphics[width=1.0\linewidth]{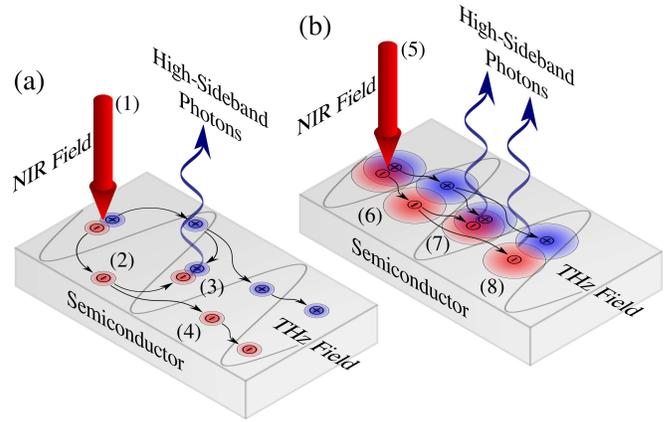}
\caption{(Color online) A schematic diagram of high-sideband generation. (a) shows the semi-classical regime. Here, the NIR field excites an electron-hole pair (1). The electron and hole are driven apart by a strong THz field such that their wavefunctions fully separate (2). Closed trajectories (3) bring the electron and hole to the same location such that they can recombine, resulting in HSG. Open trajectories (4) do not produce sidebands. (b) shows the quantum regime. Again, the NIR field excites an electron-hole pair (5). In this situation the electron-hole pair does not fully separate (6). Since there is non-zero overlap of the wavefunctions both closed trajectories (7) and open trajectories (8) can produce sidebands.} 
\label{scheme}
\end{figure}

It was predicted \cite{liu1} and subsequently observed \cite{liu2} that the related process of high-sideband generation (HSG) can occur in semiconductors. In HSG, a weak near-infrared (NIR) field promotes a valence band electron to the conduction band, creating an electron-hole pair. The pair is then driven in the bands by an intense THz frequency field. The recombination of the electron-hole pair results in the characteristic `plateau-like' spectrum with sidebands occurring at even harmonics of the THz-field and a cutoff at $3.17U_{p}$ above the bandedge [see Fig. \ref{scheme}(a)]. As with HHG, HSG has a wide range of potential applications - the solid state setting and picosecond timescales of HSG lends itself to ultrafast electro-optical applications such as Tbit/s modulation for optical communications.

The `three-step' approach to HHG (or HSG) is semi-classical. Here, the quantum evolution of the electron-ion or electron-hole pair is captured by a classical, closed trajectory that satisfies the least action principle (or the stationary phase condition in the path integral formalism)~\cite{lewenstein}. This semi-classical approximation is well justified for HHG in atoms since the excursion of the ionised electron from the atom can be as much as $50-60$ Bohr radii \cite{ivanov} and hence the overlap of the quantum wavefunctions of the driven electron and the ion ground state is minimal. 

In HSG, however, the small reduced mass of the electron-hole pair leads to an excitonic Bohr radius that can be as much as two orders of magnitude larger than the Bohr radius of an atom. In light of this, one could envisage a situation where the excursion of the electron-hole pair is on the order of the width of their wavefunctions and, hence, one would expect significant wavefunction overlap. In such a situation quantum effects would come into play and open trajectories, where the electron and hole do not return to the same location, would give a significant contribution to the sideband spectrum [see Fig. \ref{scheme}(b)]. Thus, the semi-classical description, which neglects these open trajectories, would be insufficient, motivating one to look beyond the traditional approach and to treat the process in a fully quantum mechanical way. 

Here we consider an effective mass model of a parabolic band material driven by an intense, continuous wave THz-field, $\tilde{\mathbf{F}}_{THz}(t) = \mathbf{F}_{THz}\cos\left(\omega_{THz}t\right)$, which is linearly polarized in the $z$ direction. The exciting NIR-field is also taken to be a continuous wave, $\tilde{\mathbf{F}}_{NIR}(t) = \mathbf{F}_{NIR}\mathrm{exp}\left(-i\omega_{NIR}t\right)$ with frequency, $\omega_{NIR}$, resonant with the gap. The coupling of the THz-field to the material is introduced via the time-dependent canonical momentum by the inclusion of the vector potential $\mathbf{k}\rightarrow\tilde{\mathbf{k}}(t) = \mathbf{k}+e\mathbf{A}(t)$ with $\mathbf{F} = -\partial\mathbf{A}(t)/\partial t$. Thus, for our system $\tilde{\mathbf{k}}(t) = (k_{x},k_{y},k_{z} - \frac{e}{\hbar}\int dt'\,\tilde{F}_{THz}(t'))=(k_{x},k_{y},k_{z} - \alpha\sin\left[\omega_{THz}t\right])$, with $\alpha = eF_{THz}/\hbar\omega_{THz}$ the maximum momentum obtainable from the driving field, and instantaneous band energies given by
\begin{align}
E_{eh}[\tilde{\mathbf{k}}(t)] &= E_{g}+\frac{\hbar^{2}\tilde{\mathbf{k}}(t)^{2}}{2m_{R}}\nonumber\\
 &= E_{g}+2U_{p}\left[\left(\frac{k_{z}}{\alpha}-\sin\left[\omega_{THz}t\right]\right)^{2} +\frac{k^{2}_{x}+k^{2}_{y}}{\alpha^{2}}\right].
\end{align}
Here, $E_{g}$ is the gap energy, $m_{R}$, the electron-hole reduced mass and $U_{p}=e^2F_{THz}^2/4m_{R}\omega_{THz}^2$, the ponderomotive energy. If the NIR field is sufficiently weak we can work in the single active electron-hole pair approximation. In this regime the evolution of the electron-hole pair wavefunction, $P_{\lambda}(\tilde{\mathbf{k}}, t) = \langle \hat{e}_{\lambda}^{\dagger}(\tilde{\mathbf{k}}, t)\hat{h}_{\lambda}^{\dagger}(\tilde{\mathbf{k}}, t) \rangle$ is given by
\begin{multline}
i\hbar\frac{\partial}{\partial t}P_{\lambda}(\tilde{\mathbf{k}}, t) =\left\{E_{eh}[\tilde{\mathbf{k}}(t)]-\hbar\omega_{NIR} -i\Gamma\right\}P_{\lambda}(\tilde{\mathbf{k}}, t)\\
 + i\mathbf{d}_{eh}\cdot\mathbf{F}_{NIR},
 \label{ODE}
\end{multline}
where $\lambda$ is the spin index, $\mathbf{d}_{eh} = \langle\psi_{c}|\hat{\mathbf{d}}|\psi_{v}\rangle$ is the interband dipole moment, here treated as a constant, with $|\psi_{c/v}\rangle$ denoting the Bloch states of the conduction and valence bands, respectively. $\Gamma$ is a phenomenological dephasing rate that accounts for various interband relaxation mechanisms and results in a coherence time for the electron-hole pair wavefunction of $\hbar/\Gamma$. We have not included the Coulomb interaction since the NIR field is resonant with the bandedge and hence off-resonant with the bound state excitations. We also note that the Coulomb interaction does not significantly change the maximum energy that the electron-hole pair can gain from the THz field; its effect is mainly to enhance the overall intensity of the sidebands \cite{yan}. Thus, the location of the plateau cutoffs, which are the main result of this study, should not be significantly altered by the presence of the Coulomb interaction. The output field can be found from the expectation values of the dipole moment operator
\begin{equation}
\mathbf{P}(t) = i\langle\hat{\mathbf{d}}(t)\rangle = i\sum_{\lambda}\int d^{3}k\,\mathbf{d}^{\ast}_{eh}P_{\lambda}(\tilde{\mathbf{k}}, t),
\label{Pol}
\end{equation}
from which the HSG spectrum can be calculated directly via the Fourier transform.

A full quantum mechanical treatment of the system, which takes into account open trajectories, reveals two new features that are not resolved by the semi-classical approach. Firstly, a second, quantum coherence induced plateau with a cutoff at $8U_{p}$ is observed and, secondly, this new plateau only appears when the coherence time is longer than half the THz field period, $\eta \equiv (T_{THz}/2)/(\hbar/\Gamma)\lesssim 1$.
 
The quantum coherence induced plateau cutoff at $8U_{p}$ can be understood by comparison with the method that gives the semiclassical cutoff at $3.17U_{p}$ \cite{lewenstein}. An expression for the electron-hole pair wavefunction can be found by formally integrating Eq.~\eqref{ODE}. This leads to
\begin{equation}
P_{\lambda}(\tilde{\mathbf{k}}, t) = -\frac{1}{\hbar}\int_{0}^{\infty} d\tau\,e^{iS(\tilde{\mathbf{k}},t,\tau)-i\omega t}\,\mathbf{d}_{eh}\cdot\mathbf{F}_{NIR},
\label{eh}
\end{equation}
with the semi-classical action given by
\begin{equation}
S(\tilde{\mathbf{k}},t,\tau) = -\frac{1}{\hbar}\int_{t-\tau}^{t}dt''E_{eh}[\tilde{\mathbf{k}}(t'')]-\omega_{NIR}(t-\tau)+i\frac{\Gamma}{\hbar}\tau+\omega t.
\label{action}
\end{equation}
Substituting the above result into Eq.~\eqref{Pol} leads to an expression for the polarization in terms of a path-integral-like propagator \cite{liu1, yan}
\begin{equation}
\mathbf{P}(\omega) = -\frac{i}{\pi\hbar}\int d^{3}k\int dt\int_{0}^{\infty} d\tau\,\mathbf{d}^{\ast}_{eh}e^{iS(\tilde{\mathbf{k}},t,\tau)}\mathbf{d}_{eh}\cdot\mathbf{F}_{NIR}.
\label{P}
\end{equation}
Minimizing the action in Eq. \eqref{action} leads to the classical equations of motion,
\begin{subequations}
\begin{gather}
-\frac{1}{\hbar}\int_{t-\tau}^{t}dt''\,\bm{\nabla}_{\mathbf{k}}E_{eh}[\tilde{\mathbf{k}}(t'')] = 0,\label{c1}\\
E_{eh}[\tilde{\mathbf{k}}(t-\tau)] = \hbar\omega_{NIR} - i\Gamma, \label{c2}\\
E_{eh}[\tilde{\mathbf{k}}(t)] - E_{eh}[\tilde{\mathbf{k}}(t-\tau)] = \hbar\omega_{S},\label{c3}
\end{gather}
\end{subequations}
where $\hbar\omega_{S}=\hbar\omega-\hbar\omega_{NIR}$ is the sideband energy relative to the excitation frequency. Equation \eqref{c1} states that the electron-hole pair must be at the same location to recombine (i.e. the trajectories must be closed); Eq.~\eqref{c2} that the electron-hole pairs are created at time $t-\tau$ with an energy equal to the NIR field energy and Eq.~\eqref{c3} that the sideband energy is given by the energy of the electron-hole pair at time $t$.

The semi-classical method involves simultaneously solving Eqs. \eqref{c1}-\eqref{c3} and leads to a maximum sideband energy of $3.17U_{P}$ \cite{lewenstein}. However, if the electron-hole separation is small relative to the wavefunction width there will be non-negligible overlap and hence the electron and hole do not need to exactly coincide to recombine. As a result, open trajectories can give a significant contribution and Eq. \eqref{c1} proves to be too restrictive a condition. Thus we relax the constraint in Eq. \eqref{c1}. Solving the remaining two constraint equations for an excitation resonant with the gap ($E_{g} = \hbar\omega_{NIR}$ and  $k_{x}=k_{y}=0$), we find that the real part contribution to the sidebands reads
\begin{equation}
\hbar\omega_{S} = 2U_{p}\left(\sin\left[\omega_{THz}(t-\tau)\right]-\sin\left[\omega_{THz}t\right]\right)^{2}.
\label{max}
\end{equation}
The highest sideband occurs for $\omega_{THz}t=3\pi/2$ and $\omega_{THz}\tau=\pi$, which results in a new cutoff at $8U_{P}$. Thus, the maximum sideband is given by electron-hole pairs that are created at the point of zero THz field and accelerated for a full half cycle, recombining before the THz field changes sign. Note that electron-hole pairs that travel for a time greater than $\omega_{THz}\tau=\pi$ will be decelerated and hence achieve a final energy that is less than $8U_{p}$.

The condition for the appearance of the second plateau, $\eta \equiv (T_{THz}/2)/(\hbar/\Gamma)\lesssim 1$, can be found by considering the ratio of the maximum classical excursion to the wavefunction width. The classical excursion can be found from Eq. \eqref{c1} and \eqref{c2}. For trajectories that maximizes Eq.~\eqref{max} ($\omega_{THz}t=3\pi/2, \omega_{THz}\tau=\pi$) and an NIR field that is resonant with the gap ($E_{g} = \hbar\omega_{NIR}$ and $k_{x}=k_{y}=0$), the maximum classical excursion is
\begin{equation}
\Delta x_{cl} = \frac{4\pi U_{p}}{\hbar\omega_{THz}\alpha} = \frac{e\pi F_{THz}}{m_{R}\omega^{2}_{THz}}.
\label{exc}
\end{equation}
The width of the wavefunction can be found by evaluating the integral expression in Eq. \eqref{eh}. From Eq. \eqref{max} one can see that the electron-hole pairs that give the highest sidebands are those that are excited at $\omega_{THz}(t-\tau) = \pi/2$. Thus, in order to obtain an estimate of the wavefunction width, one can expand the trigonometric functions to first order in $t-\tau$ about this point. Essentially, this approximation neglects all but the highest energy electron-hole pair trajectories. Evaluating the $\tau$ integral leads to a `Lorentzian-like' wavefunction
\begin{equation}
P_{\lambda}(\tilde{\mathbf{k}}, t) \approx \frac{e^{iS(k,t)-i\omega t}\mathbf{d}_{eh}\cdot\mathbf{F}_{NIR}}{2U_{p}\left(\frac{k^{2}}{\alpha^{2}}+2\frac{k}{\alpha}\right)-\hbar\omega_{NIR}+E_{g}-i\Gamma}.
\label{eh2}
\end{equation}
For an NIR excitation resonant with the gap one finds the peaks of the wavefunction are located at $k_{0} = 0, -2\alpha$
and the half maximum at $k_{\frac{1}{2}} = -\alpha\left(1\pm\sqrt{1 + \Gamma/2U_{p}}\right)$. If $\Gamma \ll U_{p}$, which implies that the energy gained from the THz field is much larger than the dephasing rate, we obtain a full-width-half-maximum of
\begin{equation}
\Delta k_{FWHM} = \frac{\alpha\Gamma}{2U_{p}} = \frac{2m_{R}\omega_{THz}\Gamma}{e\hbar F_{THz}}.
\end{equation}
Thus, the ratio of the classical excursion and the wavefunction width is given by
\begin{equation}
\eta \equiv \frac{\Delta x_{cl}}{\Delta x_{width}} = \Delta x_{cl}\frac{\Delta k_{FWHM}}{2} = \frac{\pi\Gamma}{\hbar\omega_{THz}} = \frac{T_{THz}}{2}\frac{\Gamma}{\hbar}.\label{eta}
\end{equation}
Equation \eqref{eta} indicates that in order to obtain sufficient overlap between the electron and hole wavefunctions after excursion, the coherence time, $\hbar/\Gamma$, should be longer than half the THz field period, i.e. longer than the duration of the quantum trajectory.

The results of the study were demonstrated numerically. Solutions to Eq. \eqref{ODE} were obtained via a standard time-domain finite-difference simulation, performed on cylindrical mesh with the $k_{z}$-axis of the grid aligned along the polarization direction the THz field \cite{liu3}. The output field, $\mathbf{P}(t)$, was calculated using Eq.~\eqref{Pol} with the corresponding HSG spectrum, $\mathbf{P}(\omega)$, found directly using a Fast Fourier transform method.

\begin{figure*}
\centering
\includegraphics[width=1.0\linewidth]{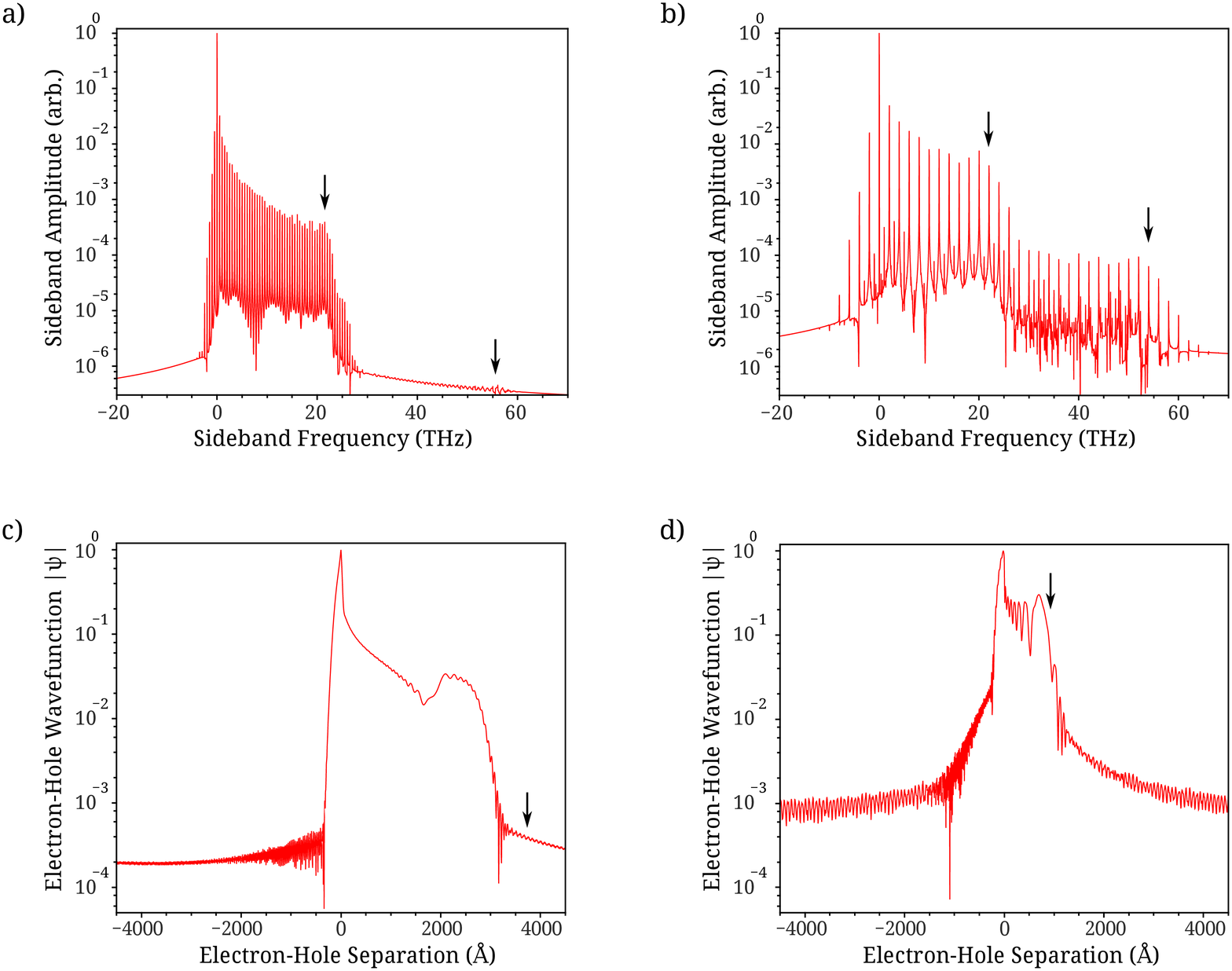}
\caption{(Color online) (a) and (b) show sideband spectra for the classical and quantum regimes. The ponderomotive energy for both cases is $U_{p} = 6.94\,THz$ and the THz field frequencies are $0.25\,THz$ in (a) and $1.00\,THz$ in (b), which correspond to THz field amplitudes of $3.06\,kVcm^{-1}$ and $12.25\,kVcm^{-1}$, respectively. The excursion-width ratios are (a) $\eta = 2.13$ and (b) $\eta = 0.53$. The arrows mark the first and second plateau cutoffs at $3.17U_{p}$ and $8U_{p}$. (c) and (d) show the the electron-hole wavefunctions as a function of separation $|\psi(x=0,y=0,z)|$ at the instance of maximal driving field for the same driving parameters as in (a) and (b), respectively. The arrow marks the maximum classical excursion of the electron-hole pair, $\Delta x_{cl}$, at the instance of maximal field.} 
\label{comp}
\end{figure*}

Figures \ref{comp} (a) and (b) show the sideband spectrum, $\mathbf{P}(\omega)$, for $U_{p} = 6.94\,THz$ and driving frequencies of $0.25\,THz$ and $1.00\,THz$ (which correspond to driving field amplitudes of $3.06\,kVcm^{-1}$ and $12.25\,kVcm^{-1}$), respectively. Material parameters were chosen to reflect GaAs ($m_{e}=0.067m_{0}$, $m_{h}=0.45m_{0}$, $E_{g}=1.424\,eV$) with $\Gamma$ taken to be $0.7\,meV$. In both cases one can see the `plateau-like' structure which is characteristic of non-perturbative processes. The two arrows on the diagrams mark the theoretical cutoffs at $3.17U_{P}$ and $8U_{P}$. For $\omega_{THz}/2\pi = 0.25\,THz$ a single cutoff occurs at $3.17U_{P}$. The excursion-width ratio in this case is $\eta = 2.13$, which implies we are in the semi-classical regime and the observed single plateau is consistent with the semi-classical theory. For $\omega_{THz}/2\pi = 1.00\,THz$ a second plateau occurs at $8U_{P}$. Here, the excursion-width ratio is $\eta = 0.53$, which implies we are in the quantum regime and the expected second, quantum coherence induced plateau is observed.

To highlight the difference between the semi-classical and quantum regimes one can compare the classical excursion of the electron-hole pair to the width of the electron-hole wavefunction. The classical excursion can be found from Eq.~\eqref{exc}. The width of the real space electron-hole wavefunction, $\psi(\mathbf{r})$, can be found from the Fourier transform of the momentum space electron-hole pair wavefunction, $P_{\lambda}(\tilde{\mathbf{k}}, t)$, and is related to the dephasing rate, $\Gamma$. Dephasing results in a finite wavefunction width; the smaller the dephasing rate, $\Gamma$, the narrower the distribution of the momentum space pair wavefunction $P_\lambda(\tilde{\mathbf{k}},t)$ and the broader the real space wavefunction, $\psi(\mathbf{r},t)$. In particular, for vanishing $\Gamma$, the momentum space electron-hole pair wavefunction will approach a $\delta$-function and, hence, be infinitely narrow. Correspondingly, the real space electron and hole wavefunction will approach plane-waves and, hence, be infinitely extended. Figure \ref{comp} (c) and (d) give $|\psi(\mathbf{r},t)|$ as a function of the electron-hole separation at the point of maximum field for the same driving field parameters as in Fig. \ref{comp} (a) and (b), respectively. The arrows mark the maximum classical electron-hole excursion, $\Delta x_{cl}$, which occurs at maximal field. For the case of the lower THz field frequency [Fig 2(c)] the overlap of the wavefunction at the classical excursion distance is minimal. Thus electron-hole pairs that follow open trajectories cannot recombine. Therefore, the second plateau is not observed. For the case of the higher THz field frequency [Fig 2(d)] there is significant overlap at the classical excursion distance. In this case electron-hole pairs need not be at the same location to recombine and, hence, open trajectories can contribute significantly to the sideband spectrum. In this case the quantum coherence induced second plateau is observed.

\begin{figure}[t]
\centering
\includegraphics[width=1.0\linewidth]{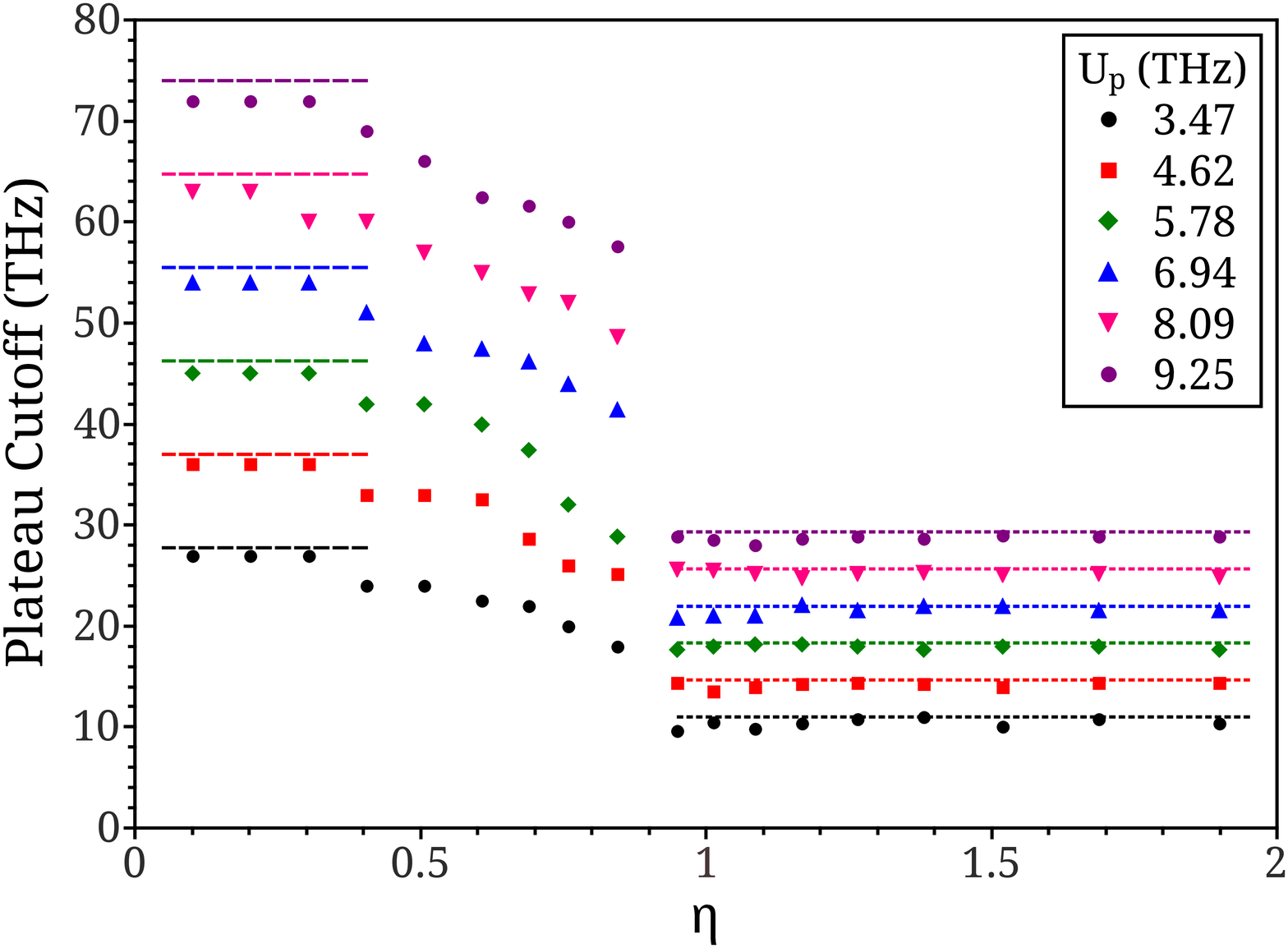}
\caption{(Color online) The plateau cutoff as a function of $\eta$, the ratio of the classical excursion to the electron-hole coherence length, for different values of the ponderomotive energy. The horizontal dotted lines to the right indicate $3.17U_{p}$. The horizontal dashed lines to the left indicate $8U_{p}$.} 
\label{fig:cutoff}
\end{figure}
Figure \ref{fig:cutoff} shows the location of the highest observed cutoff as a function of the excursion-width ratio, $\eta$. One sees the appearance of a second plateau when the classical excursion is reduced below the width of the quantum wavefunction ($\eta \lesssim 1$). 

Although we have only considered systems with parabolic bands and a direct bandgap, the results apply to a much broader range of materials. However, the effect is strongly dependent on two properties of the system, namely, the energy gain and the electron-hole coherence time. As the energy transfer is proportional to the particle velocity, materials that exhibit low effective masses or excitation conditions where the electron-hole pairs are created away from the band minima will lead to broader plateaus. Conversely, high scattering rates will reduce the coherence time and suppress the second plateau. Suitable materials need to balance both of these aspects.

We have shown that a second plateau arises owing to the quantum coherence of the optically excited electron-hole wavefunction in semiconductors when the coherence time of the wavefunction is longer than the excursion time (half the THz field period). This quantum coherence induced plateau is broader ($8U_{p}$) than the semiclassical plateau ($3.17U_{p}$), indicating sharper optical modulation in the time domain, and is determined by the maximum energy that an electron-hole pair can gain along an open trajectory. This discovery shows that HSG can occur in situations where closed trajectories do not exist. This greatly increases the range of materials (such as biased bi-layer graphene \cite{graphene}) and excitation conditions (such as lasers tuned well above band edges \cite{xt}) under which HSG can occur thereby facilitating many novel ultrafast electro-optical applications.

This work was supported by Hong Kong RGC/GRF Project 401512 and CUHK Focused Investments Scheme.

%%%%%%%%%%%%%%%%%%%%%%%%%%%%%%%%%%%%%%%%%%%%%%%%%%%%%%%%%%%%%%%%%%%%%%

\end{document}